\documentstyle[12pt,epsfig]{article}
\begin{document}
\def\d{\partial}
\def\D{\Delta}
\def\cD{{\cal D}}
\def\cK{{\cal K}}
\def\f{\varphi}
\def\g{\gamma}
\def\G{\Gamma}
\def\l{\lambda}
\def\L{\Lambda}
\def\M{{\Cal M}}
\def\m{\mu}
\def\n{\nu}
\def\p{\psi}
\def\q{\b q}
\def\r{\rho}
\def\t{\tau}
\def\x{\phi}
\def\X{\~\xi}
\def\~{\widetilde}
\def\h{\eta}
\def\bZ{\bar Z}
\def\cY{\bar Y}
\def\bY3{\bar Y_{,3}}
\def\Y3{Y_{,3}}
\def\z{\zeta}
\def\Z{{\b\zeta}}
\def\Y{{\bar Y}}
\def\cZ{{\bar Z}}
\def\`{\dot}
\def\be{\begin{equation}}
\def\ee{\end{equation}}
\def\bea{\begin{eqnarray}}
\def\eea{\end{eqnarray}}
\def\half{\frac{1}{2}}
\def\fn{\footnote}
\def\bh{black hole \ }
\def\cL{{\cal L}}
\def\cH{{\cal H}}
\def\cF{{\cal F}}
\def\cP{{\cal P}}
\def\cM{{\cal M}}
\def\olam{\stackrel{\circ}{\lambda}}
\def\oX{\stackrel{\circ}{X}}
\def\const{{\rm const.\ }}
\def\ik{ik}
\def\mn{{\mu\nu}}
\def\a{\alpha}

 \title{\sc leading role of gravity in the
structure of spinning particle}
\author{ Alexander Burinskii\\
\\
{\it Gravity Research Group, NSI
 Russian Academy of Sciences,} \\
{\it B.Tulskaye 52, Moscow 115191, Russia}}

\date{7 November, 2005}

\maketitle

 \begin{abstract}
The Kerr-Newman solution has g=2 as that of the Dirac
electron and is considered as a model of spinning
particle in general relativity.  The Kerr geometry
changes cardinally our representations on the role of gravity in
the particle physics. We show that the Kerr gravitational field
has a stringy local action and a topological peculiarity which are
extended up to the Compton distances, and also a strong non-local
action playing the key role in the mass-renormalization and
regularization of singularities.
The Kerr-Newman gravity determines the structure of spinning particle
in the form of a relativistically rotating disk, a highly oblate bag of
the Compton radius.
Interior of this bag consists of an AdS or dS ``false vacuum'',
depending on the correlation of the mass density and charge.
In the same time, the local
action of gravitational field  may be considered as
negligible for regularized particle.  \end{abstract}

\medskip


\newpage

 \section{Introduction}
The used in QED mass renormalization
is universally recognized due to an incredible exactness of
its predictions. Although its origin
lies in the classical theory of a pointlike electron, there are serious
problems with physical interpretation and mathematical correctness
of this procedure.

In this paper we consider gravitational field of spinning particle
 - the Kerr-Newman solution of the Einstein-Maxwell theory.
This solution has double gyromagnetic ratio,
as that of the Dirac electron and may be considered as a model of electron
in general relativity \cite{Car,DKS,Bur,Lop,BurBag,BurDir}.
We show that the mass renormalization and
regularization of the singularities in the Kerr-Newman source are
realized by gravitational field in a very natural manner.
It allows one to conjecture that the methodological problems of
QED may be connected with the ignorance of gravity.
QED ignores gravitational field arguing that
its local action is negligible on the scale of elementary particles.
It is true, but only partially. The unusual structure of the Kerr
solution gives us a contr-example to this assertion.
Taking the parameters of the Kerr-Newman source, charge $e$, mass $m$
and spin $J$ equal to parameters of elementary particles,
one obtains that the Kerr parameter $a=J/m,$ which characterizes
the radius of the Kerr singular ring, satisfies condition $a>>m$,
when the Kerr's event horizon disappears, and the source represents
a {\it naked singular ring of the Compton radius} $a\sim \frac {\hbar}m .$
We show that the Kerr gravity displays in the Compton the strong local
field having a stringy structure, a nontrivial topological peculiarity and
has a strong non-local action playing the key role in the mass-
renormalization. All these effects must not be ignored in a consistent
theory.

\subsection{Local and topological peculiarities of the Kerr gravity on the
Compton distances}

Let us first recall some peculiarities of the Kerr geometry.
The local action of the Kerr gravity extends on the Compton
distances due to stringy structure of the Kerr source. It was shown
\cite{BurStr,BurOri} that the ringlike Kerr singularity is indeed
a string resembling the heterotic string of superstring theory.
It breaks drastically all the former arguments based on the
assumptions on the spherical symmetry of gravitational field.
Indeed, the changes are  still more serious, since this
singular ring is a branch line of the Kerr space on two sheets,
``positive''($r>0$)  and ``negative'' ($r<0$), where $r$ is an
oblate radial coordinate of the Kerr oblate coordinate system.
Therefore, the
Kerr space has twofold topology just in this Compton region.
So, this string is ``Alice'' one \cite{Ali}, and the Kerr ring
represents a ``mirror gates'' in the ``Alice mirror world'' where
the mass and charge have another signs and fields have different
directions. It means that the behavior of the particles and fields
has to suffer from essential influence in the vicinity of this
Compton region. Note, that it is just the region which identified
in QED as a region of virtual photons.

Another very important effect of gravity is related to its
non-local action. In the next section we show that gravity provides the mass
renormalization in a very natural manner.
A smooth regularization of the
Kerr-Newman solution is considered, leading to a source in the form
of a rotating bag filled by a false vacuum. It is shown that gravity
controls the phase transition to AdS or dS false vacuum state inside
the bag, providing the mass balance.

 \section{Renormalization by gravity.}
Mass of an isolated source
is determined by asymptotic gravitational field only, and
therefore, it
depends only on the mass parameter $m$ which survives in the
asymptotic expansion for metric.
On the other hand, the total mass can be calculated as a volume integral
which takes into account densities of the electromagnetic energy
$\rho_{em},$ material (``mechanical'' mass) sources $\rho_{m}$
and energy of gravitational field $\rho_{g}$.
The last term is not taken into account in QED, but namely this
term provides renormalization.
For a spherically symmetric system, the expression may be reduced to an
integral over radial distance $r$
\be m = 4\pi \int_0^\infty \rho_{em} dr
+ 4\pi\int_0^\infty \rho_{m} dr + 4\pi\int_0^\infty \rho_{g} dr
\label{mtot} .
\ee
It looks like the expressions in
a flat space-time.
However, in the Kerr-Schild background it is consequence of the exact Tolman
relations taking into account energy of matter, energy of gravitational
field (including the contribution from pressure) and rotation  \cite{BEHM}.
In the well known classical model of electron  as a charged
sphere with electromagnetic radius $r_e =\frac {e^2}
{2m},$  integration in (\ref{mtot}) is performed in the
diapason $[r_0, \infty],$ where $r_0=r_e.$ The total mass is
determined by electromagnetic contribution only, and  contribution
 from gravity turns out to be zero. However, if
$r_0<r_e,$ electromagnetic  contribution exceeds the total mass and
this exceeding is to compensated by the negative gravitational contribution.
Indeed, the results will not depend on the cut parameter $r_0$ and,
moreover, on radial distribution of matter at all.
 Some of the terms may  be divergent, but the total
result will not be changed, since divergences will always be
compensated
by contribution from gravitational term.
It shows that, due to the strong non-local action,
gravity has to be very essential for elementary particles,
on the distances which are very far from the considered usually
Planck scale.

 \subsection{Some more on the structure of the Kerr geometry.}
 The Schwarzschild's singular point
turns in the Kerr solution into a singular ring of the radius
$a=J/m$.
For $J\sim \hbar$, it is the Compton radius which exceeds for electron
the Schwarzschild one  at $\sim 10^{22}$ times.
Angular momentum $J=\hbar /2$ for parameters of
electron is very high, $a>>m$, so the black hole horizons
disappear and the source of the Kerr spinning particle represents
a naked singular ring which may have some stringy excitations generating the spin and
mass of the extended particle  - ``microgeon'' model \cite{Bur}.
The Kerr singular ring cannot be
localized inside the region which is smaller of the Compton size,
 which is similar to the properties of the Dirac electron and to
 the assertions of QED.
One more remarkable structure of the Kerr geometry is PNC (principal
null congruence). It is  a very important
object, since the tangent to this congruence vector $k^\m$ determines
the Kerr-Schild ansatz for the metric
\be g^\mn =\eta^\mn + 2H k^\m k^n \label{KS} \ee (where $\eta^\mn$
is the auxiliary Minkowski metric) and also
the form of vector potential $ A_\m = {\cal A} (x) k_\m $
for electrically charged solution, i.e. it determines polarization of
the gravitational and
electromagnetic fields around the Kerr source.
For the nonstationary solutions containing the wave excitations
\cite{BurAxi}, PNC determines the directions
of radiation which propagates along $k^\m$.

The Kerr congruence is  determined by
the Kerr theorem in terms of twistors \cite{BurNst,Multiks} which
form a vortex of lightlike rays, see Fig.~1.
One can see, that the Kerr congruence is ``in'' -going on the
``negative''
sheet of the Kerr geometry, it passes through the disk spanned by
the Kerr ring, and turns out to be ``out''-going on the ``positive''
sheet of the space.

\begin{figure}[ht]
\centerline{\epsfig{figure=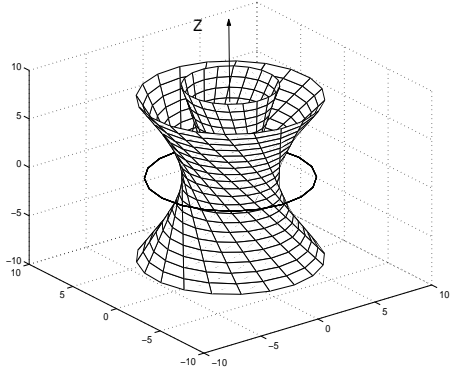,height=5cm,width=6cm}}
\caption{The Kerr singular ring and 3-D section of the Kerr
principal null congruence (PNC). Singular ring is a branch line of space,
and PNC
propagates from ``negative'' sheet of the Kerr space to ``positive '' one,
covering the space-time twice.} \end{figure}

Note, that twistor may be considered as a pair $\{x^\m,k^/m \}$ which
connect a lightlike direction, determined by the null vector
$k^\m , $ with a space-time point $x^\m$, forming a null ray
passing via a given point. Instead of the null vector $k^\m$, the
Kerr theorem treats the projective spinor coordinate $Y=\psi^0/\psi^1$
which allows to restore the null vector.

Twistorial structure of PNC forms a vortex {\it web} which covers
the whole space-time, but it focuses on the Kerr singular ring
of Compton radius. We will argue in the next section that PNC
is a flow of virtual photons (zero point field) which are fall
on the particle, excite it, and leave it out-going to infinity.

\subsection{Regularization of the Kerr singularity.}

When considering the Kerr-Newman solution as a model of electron,
one can also assume some other elementary particles.
The Kerr-Schild form of metric allows one to
consider a broad class of regularized solutions which
remove the Kerr singular ring, covering it by a matter
source.
Usually, the regularized solutions have to retain
the Kerr-Schild form of metric
and the form of Kerr principal null congruence $k_\m(x),$
as well as its property
to be geodesic and shear-free.
The space part $\vec n$ of the Kerr congruence $k_\m=(1,\vec n)$
has the form of a spinning hedgehog.  Indeed, by setting the parameter
of rotation $a$ equal to zero, the Kerr singular ring shrinks to
singular point, and $\vec n$ takes the usual hedgehog form
which is used as an ansatz for the solitonic
models of elementary particles and quarks.
It suggests that the Kerr spinning particle may have relation not only to
electron, but also to the other
elementary particles. Indeed, the Kerr-Schild class
of metric has a remarkable property, allowing us to consider a broad class of
the charged and uncharged, the spinning and spinless solutions from an unified
point of view.

The {\it smooth} regularized sources for the rotating and
non-rotating solutions of the Kerr-Schild class have the Kerr-Schild form of
metric (\ref{KS}), where the scalar function $H$ has the general form
\cite{BurBag,BEHM}

\be
H=f(r)/(r^2 + a^2 \cos^2 \theta) \label{hf}.
\ee
For the Kerr-Newman solution function $f(r)$ has
the form
\be
f(r)\equiv f_{KN}= mr -e^2/2 \label{hKN}.
\ee

Regularized solutions have tree regions:

i) the Kerr-Newman exterior, $r>r_0 $, where $f(r)=f_{KN},$

ii) interior $r<r_0-\delta $, where $f(r) =f_{int}$ and
function $f_{int}=\alpha r^n ,$ and $n\ge 4$ to suppress
the singularity at $r=0,$ and provide the smoothness of the metric
up to the second derivatives.

iii) a narrow intermediate region $r\in [r_0-\delta, r_0]$ which
allows one to get a smooth solution interpolating between regions
i) and ii).

It is advisable to consider first the non-rotating cases, since the
rotation can later be taken into account by an easy trick. In this
case, taking $n=4$ and the parameter $\alpha=8\pi \Lambda/6 ,$ one
obtains for the source (interior) a space-time of  constant
curvature $R=-24 \alpha$ which is generated by a source with
energy density
$ \rho = \frac 1 {4\pi} (f'r -f)/\Sigma^2 , $

and tangential and radial pressures
$ p_{rad}=-\rho, \quad p_{tan}=\rho - \frac 1 {8\pi}f''/\Sigma ,$

where $\Sigma=r^2.$ It yields for the interior the stress-energy
tensor in a diagonal form

$ T_{\mn} = \frac {3\alpha} {4\pi} diag (1,-1,-1,-1), $ or
\be\rho=-p_{rad}=-p_{tan}=\frac {3\alpha} {4\pi}, \label{rho} \ee
which generates a de Sitter interior for $\alpha >0,$ anti de
Sitter interior for $\alpha <0$. If $\alpha =0, $ we have a flat
interior which corresponds to some previous classical models of
electron, in particular, to the Dirac model of a charged sphere
and to the Lopez model in the form of a
 rotating elliptic shell \cite{Lop}.

The resulting sources may be considered as the bags filled by a
special matter with positive ($\alpha>0$) or negative
($\alpha < 0$)
energy density.
The transfer from the external electro-vacuum solution to the
internal region (source) may be considered as a phase transition
from `true' to `false' vacuum in a supersymmetric $U(1) \times \tilde
U(1) $ Higgs model \cite{BurBag} based on the Witten model for
superconducting strings \cite{Wit}.
Assuming that transition region iii) is very thin, one can
consider the following graphical representation which turns out to
be very useful.
\begin{figure}[ht]
\centerline{\epsfig{figure=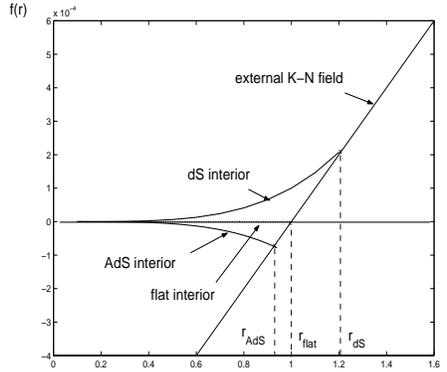,height=5cm,width=6cm}}
\caption{Regularization of the Kerr spinning particle by
matching the external field with  dS, flat or AdS interior.}
\end{figure}

The point of phase transition $r_0$ is determined by the equation
$f_{int}(r_0)=f_{KN}(r_0)$ which yields $\alpha r_0^4 = mr_0
-e^2/2 .$  We have $\rho=\frac {3\alpha} {4\pi}$
and obtain the equation

\be m= \frac {e^2} {2r_0} + \frac 4 3 \pi r_0^3 \rho . \ee

 In the first term on the right side, one can easily recognize
 the electromagnetic mass of a charged sphere with radius $r_0$,
 $M_{em}(r_0)=\frac {e^2} {2r_0}$, while the second
 term is the mass of this sphere filled by a material with a
 homogenous
 density $\rho$, $M_m =\frac 4 3 \pi r_0^3 \rho .$
 Thus, the point of intersection $r_0$ acquires
a deep physical meaning, providing the energy balance by the mass
formation. In particular, for the classical Dirac model
of a charged sphere with radius $r_0=r_e=\frac {e^2} {2m},$ the
balance equation yields the flat internal space with $\rho=0.$ If
$r_0> r_e$, a material mass of positive energy $M_m>0$ gives a
contribution to total mass $m$. If $r_0 < r_e ,$ this contribution
has to be negative $M_m <0 , $ which is accompanied by the
formation of an AdS internal space.

 \subsection{Transfer to the rotating case.}
All the above treatment retains valid for the rotating cases, and
for the passage to a rotating case, one has only to set
$ \Sigma=r^2 +a^2 \cos^2 \theta , $ and consider $r$ and
$\theta$ as the oblate spheroidal coordinates \cite{BEHM}.

The Kerr-Newman spinning particle with a spin $J=\frac 1 2 \hbar,$
 acquires the form of a relativistically rotating disk, so
the board of the disk has $v\sim c$ \cite{BEHM}.
The corresponding diagonal
stress-energy tensor describes in this case the matter
of source in a co-rotating with this disk coordinate system.
Disk has the form of a highly oblate
ellipsoid with the thickness $r_0$ and the Compton radius
$a=\frac 1 2 \hbar/m$. Interior of the disk
represents a ``false'' vacuum having superconducting properties
which are modelled by the Higgs field.
It is a physical realization of the ``Alice mirror world''
\cite{Ali}.
The charges are concentrated on the surface of this disk, at
$r=r_0$. Inside the disk the local gravitational field is
negligible.

 \section{Regularization of the zero-point radiation.}
The Kerr singular string may acquire electromagnetic wave
excitations from interaction with virtual photons
\cite{Bur,BurAxi,BurOri}. In classical theory these excitations
have to lead to a radiation and non-stationarity of solutions.
In the Kerr-Schild formalism \cite{DKS}
electromagnetic excitations and radiation are related to some
field $\gamma(x)$, and are described by the $\{13\}$-tetrad components
of the self-dual tensor
\be {\cal F}_{31}=\gamma Z +(AZ),_1  \approx \gamma \frac  1{r} +
(extra  \ terms) .\ee
It leads to an electromagnetic radiation and a flow of energy
along the Kerr congruence, $k_\m$, and consequently, to a
nonstationarity of the solutions.
Note, that in quantum theory oscillations are stationary and absence of
radiation caused by oscillations is postulated.

The treatment of quantum field theory in
curved spaces  \cite{deWit} gives a receipt for transfer from
quantum fields to the classical Einstein-Maxwell theory. It shows
that quantum fields are concentrated in the divergent vacuum zero
point field, and by the transfer to the Einstein-Maxwell equations,
 the stress-energy tensor has to be regularized by subtraction of
 the quantum vacuum fields \cite{deWit}.
\be T^{(reg)}_{\mn} = T_{\mn} - <0|T_{\mn}|0>. \ee
Regularization has to satisfy the condition
$ T^{(reg) \ \mn} ,_\m = 0 . $

It suggests \cite{BurAxi,BurOri,BurNst} that the part of
stress-energy tensor, which is related to electromagnetic
radiation propagating along the Kerr
PNC, has to be regularized by a special subtraction as a vacuum
field, before the substitution into the Einstein-Maxwell field
equations. Twofoldedness of the Kerr geometry confirms this point of view,
since {\it the out-going radiation on the ``positive'' out-sheet
of the metric is compensated by an in-going radiation
on the ``negative'' in-sheet}, see Fig.~1.
So, physically, there is no reason for the lost of mass
by radiation. It shows, that the term
${\cal F}_{31} = \gamma \frac  1{\tilde r}$ has to be
identified with the vacuum zero-point field. In this case
the electromagnetic excitations on the Kerr background
 may be interpreted as a resonance of the zero-point fluctuations
 on the (superconducting) source of the Kerr spinning particle
\cite{BurAxi,BurOri}.
It was shown \cite{BurAxi,BurDir} that such a regularization
may be performed and leads to some modified Kerr-Schild equations.
 Although, the exact nontrivial solutions of the regularized system
have not been obtained so far, there were obtained corresponding
exact solutions of the Maxwell equations which show that any
``aligned'' excitation of the Kerr geometry leads to the
appearance of extra ``axial'' singular lines (strings)
\cite{BurAxi,BurTwi}
which are semi-infinite. These strings are related to solutions of the
Dirac equation, are modulated by de Broglie periodicity and may be
considered as the physical carriers of wave function.
\section*{Conclusion. Multi-sheeted twistorial web}
 The obtained recently multiparticle Kerr-Schild
solutions \cite{Multiks} support the above point of view.
It was obtained that the Kerr theorem has a wonderful consequence:
the twistorial webs related to different particles penetrate trough each other
without interaction. The extended twistorial
space-time, consisting of the pair $\{x,Y\},$ where
$x\in M^4, \ Y \in CP^1,$ turns out to be multi-sheeted, similar
to multi-sheeted Riemann holomorphic surfaces. In the same time,
the obtained exact multiparticle solutions show, that metric turns
out to be singular along some twistor lines which are {\it common}
for twistorial structures of two interacting particles. For any two
interacting particles one can find two such common twistor lines
of opposite direction. As a result, a pair
of semi-infinite axial string appears by interaction with any
external particle, and interaction with many surrounding
particles results that almost all the twistorial rays of the Kerr
PNC turn out to be singular. It may be interpreted as an evidence
that PNC describes a virtual multiparticle interaction,
and that related to PNC fields have to be regularized as belonging
to the virtual vacuum fields.
We arrive at the conclusion that the  twistorial structure of the
Kerr PNC belongs to the vacuum zero-point field and hints at
a multi-sheeted twistorial texture of vacuum \cite{BurAxi,BurDir}.
\bigskip

\section*{Acknowledgments}  This work was supported by RFBR project
04-02-17015. Author thanks Organizers, and personally, Miroslav
Finger for very kind invitation to
attend this Conference and for financial support. Author is also
thankful to A.Efremov, S.Gerasimov and O.Teryaev
for very useful conversations.

 \end{document}